\documentclass[aps,twocolumn,nofootinbib, superscriptaddress, floatfix,longbibliography]{revtex4-1}

\usepackage{multirow}
\usepackage{booktabs}
\usepackage{placeins}
\usepackage{soul}
\usepackage{bbm}
\usepackage{amsmath,amssymb,amsbsy,bm,amsfonts,mathrsfs}
\usepackage{graphicx}
\usepackage{wrapfig}
\usepackage{url}
\usepackage{float}
\usepackage{lipsum}
\usepackage{enumitem}
\usepackage{titlesec}
\usepackage{microtype}
\usepackage{slashed}

\usepackage[usenames,dvipsnames,svgnames]{xcolor}
\definecolor{redd}{rgb}{0.8, 0.1,0.2}
\definecolor{navy}{rgb}{0.05, 0.23,0.75}

\usepackage[colorlinks]{hyperref}
\hypersetup{
    colorlinks    =    true,
    citecolor     =    navy,
	linkcolor     =    redd,
	urlcolor      =    navy,
	anchorcolor   =    blue
}

\begin{document}

\title{CDF W mass anomaly revisited }

\author{Shou-hua Zhu}
\email{shzhu@pku.edu.cn}
\affiliation{Institute of Theoretical Physics, Peking University, Beijing 100871, China}
\affiliation{Collaborative Innovation Center of Quantum Matter, Beijing, 100871, China}
\affiliation{Center for High Energy Physics, Peking University, Beijing 100871, China}

\begin{abstract}

The CDF, ATLAS and LHCb have released the measurements on the W boson mass $m_W$ at $\sqrt{S}=1.96, 7, 13 TeV$, respectively. The measured values show the declining tendency, namely $m_W$ decreases with the increment of the collider energy. If the declining tendency is confirmed, it might be the signal of metric field at high energy colliders. In this paper, we propose a model to account for such tendency and explore the properties of the model.
\end{abstract}

\maketitle

\section{Introduction}

In 2022, the new analysis of W mass based on CDF data with 8.8  $fb^{-1}$ integrated luminosity was released. CDF is one of the detector at $p\overline{p}$ collider Tevatron with energy as high as
$\sqrt{S}=1.96 TeV$. The result
\cite{CDF:2022hxs}
\begin{eqnarray}
m_W=80,433.5\pm 9.4 MeV \label{wmass-cdf}
\end{eqnarray}
is quite extraordinary since it is more than $7\sigma$ higher than
that of the standard model (SM) global fit. It has stimulated numerous studies
\cite{CDF:refs} on
the discrepancy.
As the counter part at the Tevatron, D0 result is \cite{D0:2012kms}
\begin{eqnarray}
m_W=80,375\pm 11 \pm 20 MeV. \label{wmass-d0}
\end{eqnarray}
Since CDF and D0 disagree significantly, further inputs from other  experiments are necessary.
The author has experienced many anomalies which have come and gone. One only needs to take the interesting anomaly seriously before it can be confirmed. Recently we are considering to drop the single Ricci scalar term in the Lagrangian and looking for the phenomenological evidences. The CDF new result happens to be one of the indications, which is why we wrote this paper more than one year later after CDF paper \cite{CDF:2022hxs}.

In order to explore the unknown physics from the experimental measurement, the more reliable way is to examine firstly the single experiment, at least with the same energy. After all, the different experiments with various colliding beams and energy may bring the unknown physical effect, besides the man-made mistakes. In the following we will examine other W mass measurements.

Compared with CDF analysis, W mass is measured to be \cite{ATLAS:2017rzl}
\begin{eqnarray}
m_W= 80,370\pm 19 MeV \label{wmass-atlas}
\end{eqnarray}
by ATLAS using
$\sqrt{S}=7 TeV$ data. It is one of the detectors at $pp$ collider LHC. The discrepancy between CDF and ATLAS is around $3\sigma$.
W mass is measured to be \cite{LHCb:2021bjt}
\begin{eqnarray}
m_W=80,354\pm 23\pm 22 MeV \label{wmass-lhcb}
\end{eqnarray}
by LHCb using $\sqrt{S}=13 TeV$ data.
There were also other analysis results at four detectors ALEPH, DELPHI, L3 and OPAL at $e^+e^-$ collider LEP \cite{ALEPH:2010aa}. However the central values of the old LEP results are quite diverse, and the uncertainty of any single detector is quite larger than those of CDF and ATLAS.

From Eq. (\ref{wmass-cdf},\ref{wmass-atlas}, \ref{wmass-lhcb}), we can see that the central values of $m_W$ decline with the increment of the colliding energy. Frankly speaking, the statistical significance of this declining tendency is not so high. It is quite interesting to know wether the future CMS analysis with $\sqrt{S}=7,8$ and $13 TeV$ can confirm the tendency or not.

In fact, there are similar declining tendency for top quark mass measurement.
The latest top quark mass, combined CDF and D0 data, is \cite{ParticleDataGroup:2022pth}
\begin{eqnarray}
m_t=174.30\pm 0.35 \pm 0.54 GeV \label{topmass-cdfd0}
\end{eqnarray}
by direct measurement at Tevatron with
$\sqrt{S}=1.96 TeV$.
At higher energy LHC, top quark mass is
\begin{eqnarray}
m_t&=& 172.69\pm 0.25\pm 0.41 GeV \label{tomassATLAS}\\
m_t&=&172.44\pm 0.13\pm 0.47 GeV \label{tomassCMS}
\end{eqnarray}
at ATLAS \cite{ATLAS:2018fwq} and CMS \cite{CMS:2015lbj} respectively with $\sqrt{S}=7, 8 TeV$ data.
The latest CMS result using $\sqrt{S}=13 TeV$ data is \cite{CMS:2023ebf}
\begin{eqnarray}
m_t&=& 171.77 \pm 0.37 GeV.
\end{eqnarray}
In fact, it is only one measurement and fair to wait for the combination value with other $13 TeV$ data.

One may naturally wonder how about the mass of Z boson? Actually there was the very precise measurement from
Z-pole data at LEP-1.
However, meaningful precise measurement from other experiments is absent. Although the
precision from other experiments is not comparable with that of LEP-1 \cite{ParticleDataGroup:2022pth}, it is quite interesting and important to examine above-mentioned declining tendency for $m_Z$ at the LHC.

We have enumerate tediously the measured values of heavy particles, $m_W$,$m_Z$ and $m_t$. Note that they should be the constant in the SM since
the energy scale dependence has been removed after including the higher order effects during the data fitting. Namely
their values should be the same at the different colliders: LEP, Tevatron and LHC. If the measured values are really decrease with the increment of collider energy, how to account for such effects and declining tendency? In this paper, we try to attribute such effects to the different metric field for different colliding energy. On one hand, it is quite natural since the different energy will cause different metric field at the reaction region. On the other hand, such effects would be thought negligible tiny, since they are usually suppressed by Planck scale according to the common wisdom. However, the physics of the highest energy regime reached by high energy colliders should be basically assumed as not fully known. Such possibility is still open. In the following we will focus on such theoretical explanation.

\section{The Model}

The Lagrangian of proposed model can be written as
$$
\mathcal{L}=\mathcal{L}_g+\mathcal{L}_m.
$$
Here the general coordinate invariant pure gravity and matter Lagrangians with the metric field $g$ and the weak doublet Higgs field $\Phi$ are
\begin{eqnarray}
\mathcal{L}_g= \sqrt{g}\left\{-\kappa R\right\}
\label{puregravity}
\end{eqnarray}
\begin{eqnarray}
\mathcal{L}_m &=& \sqrt{g} \left\{ - g^{\mu\nu} \partial_\mu\Phi^\dagger  \partial_\nu \Phi + \lambda \left(\Phi^\dagger \Phi\right)^2 \right. \nonumber \\
&& \left. +
\Phi^\dagger \Phi \left(\xi R - \mu^2\right) +\Lambda_0 \right\}. \label{eq1}
\end{eqnarray}
The convention is the same with Ref. \cite{tHooft:1974toh}, namely purely imaginary time coordinate with $ g_{\mu\nu}=\delta_{\mu\nu}$ for flat space, and the gravitational coupling $\frac{c^4}{16\pi G_N}\equiv\frac{m_P^2}{16\pi }$ is chosen as 1. Here $\kappa$ is a free parameter to be determined.
$\mu^2$ term is needed as in the usual SM Lagrangian, which literally
induces mass of all particles in the SM after electro-weak symmetry spontaneously breaking. $R$ is the usual Ricci scalar and $\xi$ is a dimensionless free parameter. $\Lambda_0$ is the allowed free constant parameter.

The electro-weak symmetry breaking is realized through Higgs field acquiring the
vacuum expectation value $(v)$
$$
<\Phi>=v+H
$$
and the $H$ is the physical Higgs field. Here
\begin{eqnarray}
v^2=\frac{1}{2} \frac{\mu^2-\xi R}{\lambda}
\label{eqvev}
\end{eqnarray}
Different colliders correspond to different $R$. Currently the detail calculation for high energy collisions are not available. In the long run, R should be calculated in the colliding case. As the simplest approximation, R is
 assumed to be a different constant for different energy collision, which can
 be extracted from experiment data.
 As usual, the $m_W$,$m_t$, $m_Z$ and $m_H$ are all proportional to $v$. As shown in Eq. (\ref{eqvev}), the declining tendency for different energy colliders should be universal behavior.

 Based on current mass measurements, the order of magnitude of variation among LEP, Tevatron and LHC should be $O(10^{-3})\sim O(10^{-4}) $. Can the contribution from $\xi R$ be so large? Basically $\xi$ is an arbitrary  dimensionless parameter, and which can
 be determined empirically. We will argue theoretically that the contribution can be large. Due to the renormalizable criteria, as discussed in next section, we will drop R-term in Eq. (\ref{puregravity}). In order to reproduce the usual Hilert-Einstein gravity, $\xi$ is fixed to be $O(m_P^2/v^2)$. Such value is much larger than
 the usual assumption, for example for the case of Higgs inflation models.
 Usually the range with the sizable gravity effect is estimated as
 $$
 r \sim  G_N E=\frac{E}{m_P^2},
 $$
 where $E$ is the effective collider energy.
 Due to the $\xi$ enhancement, the range becomes
 $$
 r \sim \xi \frac{E}{m_P^2}\sim \frac{m_P^2}{v^2} \frac{E}{m_P^2}=\frac{E}{v^2}.
 $$

After the electro-weak symmetry breaking with $
v^2= \mu^2/ (2 \lambda)$,
$$
-\kappa+\xi v^2= -1
$$
which is empirically required by the Newtonian gravity.
And $$
\lambda v^4 - \mu^2 v^2 + \Lambda_0=\Lambda
$$
which is the cosmological constant.
After symmetry breaking, the induced Lagrangian becomes
\begin{eqnarray}
\mathcal{L}= \sqrt{g} \left\{ 
-R +\Lambda + \cdots \right\}. \label{eq2}
\end{eqnarray}

\section{Theoretical reason to drop $R$ term}

In order to make the $\xi R$ sizable contribution at the high energy collder plausible, we will explore the theoretical motivation to drop $R$ term, namely $
\mathcal{L}_g=0$ due to renormalizable criteria.

Renormalizability and associated infinity are usually thought as annoying, however it can be treated as the tool, even a principle to construct a meaningful theory. In order to illustrate the key difficulty to renormalize gravity. We utilize a toy model with only one real scalar field $\phi$. $\kappa$ is taken as 1 in Eq. (\ref{puregravity}). The Lagrangian of matter of Eq. (\ref{eq1}) is replaced by
\begin{eqnarray}
\mathcal{L}_m= \sqrt{g}\left\{-\frac{1}{2} g^{\mu\nu} \partial_\mu \phi \partial_\nu \phi + \frac{1}{2} \phi M^2 \phi \right\}.
\label{eq3}
\end{eqnarray}

Treating $g$ as the external source, the counter-terms at one-loop level can be extracted from Ref. \cite{tHooft:1974toh}
\begin{eqnarray}
\Delta {\mathcal{L}}&=& \frac{\sqrt{g}}{\epsilon} \left\{ \frac{1}{4} \left( M^2 -\frac{1}{6} R \right)^2 \right. \nonumber \\
&& \left. +\frac{1}{120} \left( R_{\mu\nu} R^{\mu\nu} -\frac{1}{3} R^2 \right) \right\} \label{eq4}
\end{eqnarray}
where $\epsilon=8\pi^2 (D-4)$ and $D$ is the dimension of the space-time.

Ref. \cite{tHooft:1974toh} has argued that the unrenormalizable term in  Eq. (\ref{eq4}), namely the $ R M^2$ term, can be eliminated by adding the specific term $\frac{1}{12}R \phi^2$ to the original Lagrangian of Eq. (\ref{eq3}). However the unrenormalizable terms $R^2$ and $R_{\mu\nu} R^{\mu\nu}$ remain. The situation  becomes even worse after including contributions from the gravitons in the loops.  It seems impossible to generally eliminate all unrenormalizable terms by modifying the original Lagrangian. This is the key argument that gravity is an unrenormalizable theory. As shown in this simple excise, there exists fundamental difficulty to renormalize the gravity in this way. Some fundamental aspect of the gravity has to be changed.

As the basic requirement of a renormalizable theory, the new form counter-terms beyond the origin Lagrangian are not allowed. It seems there is only possible by treating the metric as the coupling parameter instead of the dynamical field. Under this assumption, the metric is not the dynamical field, namely the kinetic term $R$ will be dropped. Provided that the metric acts only as the parameter, the form of counter-terms in Eq. (\ref{eq4}) is the same with original Lagrangian in Eq. (\ref{eq3}). All the previous unrenormalizable terms $ R M^2$, $R^2$ and $R_{\mu\nu} R^{\mu\nu}$ are the functions of the $g_{\mu\nu}$, which is the building block
of the original Lagrangian. From this point of view, the model must be renormalizable as it should be. The renormalizability of toy model of Eq. (\ref{eq3}) is guaranteed by the properties of the dynamical quantum field $\phi$. The metric only becomes the dynamical field after the electro-weak symmetry breaking, as shown in last section.

In principle, a realistic model should include all theoretical allowed terms. In Eq. (\ref{puregravity}) and Eq. (\ref{eq1}),
the kinetic term $R$ and the higher power of $R$ terms are not allowed, since these terms break either the theory renormalizability or vacuum stability. In this sense, the renormalizabilty is treated as the principle to construct
a physical theory.

For the general case, the quantum behavior can be written as the path integral of
dynamical field $\phi$
\begin{eqnarray}
Z=\int D\phi \exp\left\{ i S\right\}.
\end{eqnarray}
Note that the metric field $g$ is not a priori assumed as dynamic field.
The action $S$ can be divided as metric and other (matter) parts
\begin{eqnarray}
S= S(g)+ S(g,\phi)+ S(\phi).
\end{eqnarray}

As such, $\exp \{i S(g)\}$ is independent on the quantum field ($\phi$) and can be dropped and the path integral can be simplified as
\begin{eqnarray}
Z=\int D\phi \exp\left\{ i S(g, \phi)
+i S(\phi) \right\}.
\end{eqnarray}
The metric, as the dynamical field after electro-weak symmetry breaking, manifests itself only classically. Eq. (\ref{eq1}) is only the specific realization of the general case.

\section{Conclusion and discussion}

This paper has explored the possible implication of CDF W mass anomaly. We propose a model to account for the possible collider energy dependence measurements of $m_W$ and $m_t$. If such dependence is confirmed, it may be the signature of the metric field at high energy collider. The several future experimental measurements are warmly welcomed, especially the $m_W$, $m_t$, $m_Z$ and $m_H$ at the LHC with $\sqrt{S}=7/8$ and $13 TeV$. Note that the global fit for various experiments with different energy is illegal if the metric field sizable contributions are not included. Meanwhile the model also influences the Higgs study in the high energy regime and the early evolution of the Universe.

\section*{Availability of data and material}

The data analyzed during the current study are all available from the published papers or preprints.

\section*{Competing interests}

The authors declare that they have no competing interests.

\section*{Funding}

This work is supported by the National Science Foundation of China under Grants No. 11635001, 11875072.

\section*{Authors' contributions}

Shou-hua Zhu is the sole author with all responsibility of the draft.

\section*{Acknowledgments}

The author thanks Hai-Bo Li and Qiang Li for illuminating discussions on Z mass measurement.


\begin{thebibliography}{99}

\bibitem{CDF:2022hxs}
T.~Aaltonen \textit{et al.} [CDF],
Science \textbf{376}, no.6589, 170-176 (2022)
doi:10.1126/science.abk1781

\bibitem{CDF:refs}
There are over 300 papers which can be
found via the citations of Ref. \cite{CDF:2022hxs},
for example 
M.~D.~Zheng, F.~Z.~Chen and H.~H.~Zhang,
AAPPS Bull. \textbf{33}, no.1, 16 (2023)
doi:10.1007/s43673-023-00086-3
[arXiv:2204.06541 [hep-ph]].


\bibitem{D0:2012kms}
V.~M.~Abazov \textit{et al.} [D0],
Phys. Rev. Lett. \textbf{108}, 151804 (2012)
doi:10.1103/PhysRevLett.108.151804
[arXiv:1203.0293 [hep-ex]].

\bibitem{ATLAS:2017rzl}
M.~Aaboud \textit{et al.} [ATLAS],
Eur. Phys. J. C \textbf{78}, no.2, 110 (2018)
[erratum: Eur. Phys. J. C \textbf{78}, no.11, 898 (2018)]
doi:10.1140/epjc/s10052-017-5475-4
[arXiv:1701.07240 [hep-ex]].

\bibitem{LHCb:2021bjt}
R.~Aaij \textit{et al.} [LHCb],
JHEP \textbf{01}, 036 (2022)
doi:10.1007/JHEP01(2022)036
[arXiv:2109.01113 [hep-ex]].

\bibitem{ALEPH:2010aa}
 [ALEPH, CDF, D0, DELPHI, L3, OPAL, SLD, LEP Electroweak Working Group, Tevatron Electroweak Working Group, SLD Electroweak and Heavy Flavour Groups],
[arXiv:1012.2367 [hep-ex]].

\bibitem{ParticleDataGroup:2022pth}
R.~L.~Workman \textit{et al.} [Particle Data Group],
PTEP \textbf{2022}, 083C01 (2022)
doi:10.1093/ptep/ptac097

\bibitem{ATLAS:2018fwq}
M.~Aaboud \textit{et al.} [ATLAS],
Eur. Phys. J. C \textbf{79}, no.4, 290 (2019)
doi:10.1140/epjc/s10052-019-6757-9
[arXiv:1810.01772 [hep-ex]].

\bibitem{CMS:2015lbj}
V.~Khachatryan \textit{et al.} [CMS],
Phys. Rev. D \textbf{93}, no.7, 072004 (2016)
doi:10.1103/PhysRevD.93.072004
[arXiv:1509.04044 [hep-ex]].


\bibitem{CMS:2023ebf}
 [CMS],
[arXiv:2302.01967 [hep-ex]].



\bibitem{tHooft:1974toh}
G.~'t Hooft and M.~J.~G.~Veltman,
Ann. Inst. H. Poincare Phys. Theor. A \textbf{20}, 69-94 (1974)

\end{thebibliography}
%

\end{document}